\documentclass[%
 aip,
 apl,%
 amsmath,amssymb,
 reprint,%
]{revtex4-1}

\usepackage{graphicx}% Include figure files
\usepackage{dcolumn}% Align table columns on decimal point
\usepackage{bm}% bold math
\usepackage{epstopdf}
\usepackage{romannum}
\usepackage{overpic}
\usepackage{xcolor}

\begin{document}

\preprint{AIP/123-QED}

\title[Sample title]{POLED displays: Robust printing of pixels}% Force line breaks with \\
%\thanks{Footnote to title of article.}

\author{Pallav Kant}
 \affiliation{School of Mathematics and MCND, University of Manchester, Oxford Road, Manchester M13 9PL, United Kingdom}%Lines break automatically or can be forced with \\
\author{Andrew Hazel}
 \affiliation{School of Mathematics and MCND, University of Manchester, Oxford Road, Manchester M13 9PL, United Kingdom}%Lines break automatically or can be forced with \\
\author{Mark Dowling}%
\affiliation{Cambridge Display Technology, Technology Development Centre, Unit 12, Cardinal Business Park, Godmanchester, Cambridgeshire PE29 2XG, United Kingdom%\\This line break forced with \textbackslash\textbackslash
}%
\author{Alice Thompson}
\affiliation{School of Mathematics and MCND, University of Manchester, Oxford Road, Manchester M13 9PL, United Kingdom}%Lines break automatically or can be forced with \\
\author{Anne Juel}
\email{anne.juel@manchester.ac.uk}
% \homepage{http://www.Second.institution.edu/~Charlie.Author.}
\affiliation{MCND and School of Physics and Astronomy, University of Manchester, Oxford Road, Manchester M13 9PL, United Kingdom%\\This line break forced% with \\
}%

\date{\today}% It is always \today, today,

\begin{abstract}
%Max 250 words: currently 158
The fabrication of a high-quality POLED (Polymeric Organic Light
Emitting Diode) display requires the deposition of identical, uniform
fluid films into a large number of shallow recessed regions that form
a regular array of pixels on a display backplane. 
We determine the protocols required to achieve continuous liquid coverage of
the entire pixel area for the case where
equally-spaced fluid droplets
are sequentially printed along a straight line within a stadium-shaped
pixel, and explore how these protocols depend on the wetting properties of the pixel
surface. Our investigation uses 
a combination of experiments and numerical modelling, based on the
assumption of fluid redistribution via capillary spreading according
to a Cox-Voinov law. We show that the model is able to predict quantitatively the
evolution of the liquid deposited in a pixel and
provides a computationally inexpensive design tool to determine
efficient printing strategies that account for uncertainties arising
from imperfect substrate preparation or printhead dysfunction.

\end{abstract}

\maketitle

In recent years, inkjet printing-based manufacturing\cite{calvert2001inkjet,singh2010inkjet} has established itself as a simpler and more economical approach than conventional lithography processes for the production of a variety of microelectronics,\cite{sirringhaus2000high,minemawari2011inkjet,ko2007all} displays,\cite{andersson2007printable,shimoda2003inkjet,halls2005ink} sensors,\cite{dua2010all,lorwongtragool2014novel,abe2008inkjet} photovoltaic cells\cite{hoth2008printing,jung2014all}, etc.
%This method involves deposition of a functional material onto the target substrate via a carrier liquid in the form of droplets.
However, limitations pertaining to the stability of functionalized
inks and the need for robust printing strategies that will yield controlled thin-film deposits of non-trivial shapes\cite{stringer2010formation,duineveld2003stability,dalili2014formation}, mean that the technology has had limited uptake on an industrial scale\cite{sowade2016up,noh2015key}.
In this letter, we explore printing strategies that can be used to form an elongated thin-film ($L\simeq 200~\mu$m) of specified shape by sequential deposition of partially overlapping droplets; a process which arises in the context of POLED (Polymeric Organic Light Emitting Diode) display fabrication\cite{halls2005ink,shimoda2003inkjet}.

%Fabrication of a POLED display via inkjet printing involves the deposition of a pattern of identical thin-films of electroluminescent polymeric inks in close proximity. They form the smallest light emitting units on a display backplane \cite{halls2005ink}, also known as pixels. A high density of pixels per unit area of display ensures a smooth projected image and the absence of artefacts visible to the naked eye. Therefore, the main challenge associated with the fabrication of high-resolution displays via the inkjet method is to robustly deposit precise volumes of liquid, typically a few tens of pico-litres, to achieve thin-films in an exact pixel shape within a repeating pattern.
%This is best achieved by introducing chemical\cite{wang2004dewetting,gau1999liquid,kuang2014controllable} or geometrical\cite{hendriks2008invisible,seemann2005wetting,kant} substrate patterns to manipulate the spreading of the deposited liquid, or even a combination of both\cite{kant2}.
%However, high performance POLED displays also rely on the controlled coverage of the entire pixel area by a thin liquid film in order to ensure that a uniform layer of solidified material remains upon evaporation of the solvent. 
%In this study, we combine experiments and numerical simulations of fluid deposition into a single stadium-shaped pixel used in POLED display development to determine the influence on its filling of the wetting properties of the substrate and of the operating conditions, i.e. the total number of droplets deposited in a pixel ($N$) and the inter-drop distance between them ($\Delta x$).

The manufacture of high-resolution POLED displays \textsl{via} inkjet
printing requires the controlled deposition of a dense pattern of
identical liquid films of a functional material in solution. Solid
deposits are formed upon evaporation of the solvent to yield the
bottom-gated transistors that emit light, commonly referred to as
pixels \cite{halls2005ink}. The formation of continuous solid
deposits, which is essential to achieve high-performance displays,
requires the coverage of the entire pixel area by the deposited liquid
film. In addition, an accurately defined pixel edge is required in
order to deter unwanted leakage currents and crosstalk between the
pixels \cite{singh2010inkjet}. This is a considerable challenge for
the inkjet-printing of thin-films which tends to suffer from low
spatial resolution
\cite{duineveld2003stability,soltman2008inkjet,stringer2010formation,
  alice}. Hence,
chemical\cite{wang2004dewetting,gau1999liquid,kuang2014controllable}
or geometrical\cite{hendriks2008invisible,seemann2005wetting,kant}
substrate patterns are commonly introduced to manipulate the spreading
of the deposited liquid. Kant et al. \cite{kant2} recently identified
the distinct roles of topography and wettability patterning in the
sequential deposition of partially overlapping droplets within a
shallow recessed substrate region (pixel) that was highly wetting
compared to the elevated banks surrounding it. They found that the
presence of bounding side walls enhances local spreading thus
facilitating fluid coverage of the entire recessed region, whereas
wettability patterning ensures containment of the fluid within the
pixel. Hence, \textsl{robust} filling of pixels requires both forms of patterning in combination.

In this letter, we focus on determining the conditions required to
fill such pixels as a function of the wetting properties of the
substrate and the printing parameters, i.e. the total number of
droplets deposited in a pixel ($N$) and the inter-drop distance between
them ($\Delta x$). We demonstrate that a simple model\cite{kant,kant2} based on capillary spreading of the deposited liquid provides a computationally inexpensive design-tool to determine safe printing strategies that account for experimental limitations pertaining to the patterning accuracy and printhead operation. 

\begin{figure}
\centering
\includegraphics[clip, trim=0cm 0cm 0cm 0cm, width=0.45\textwidth]{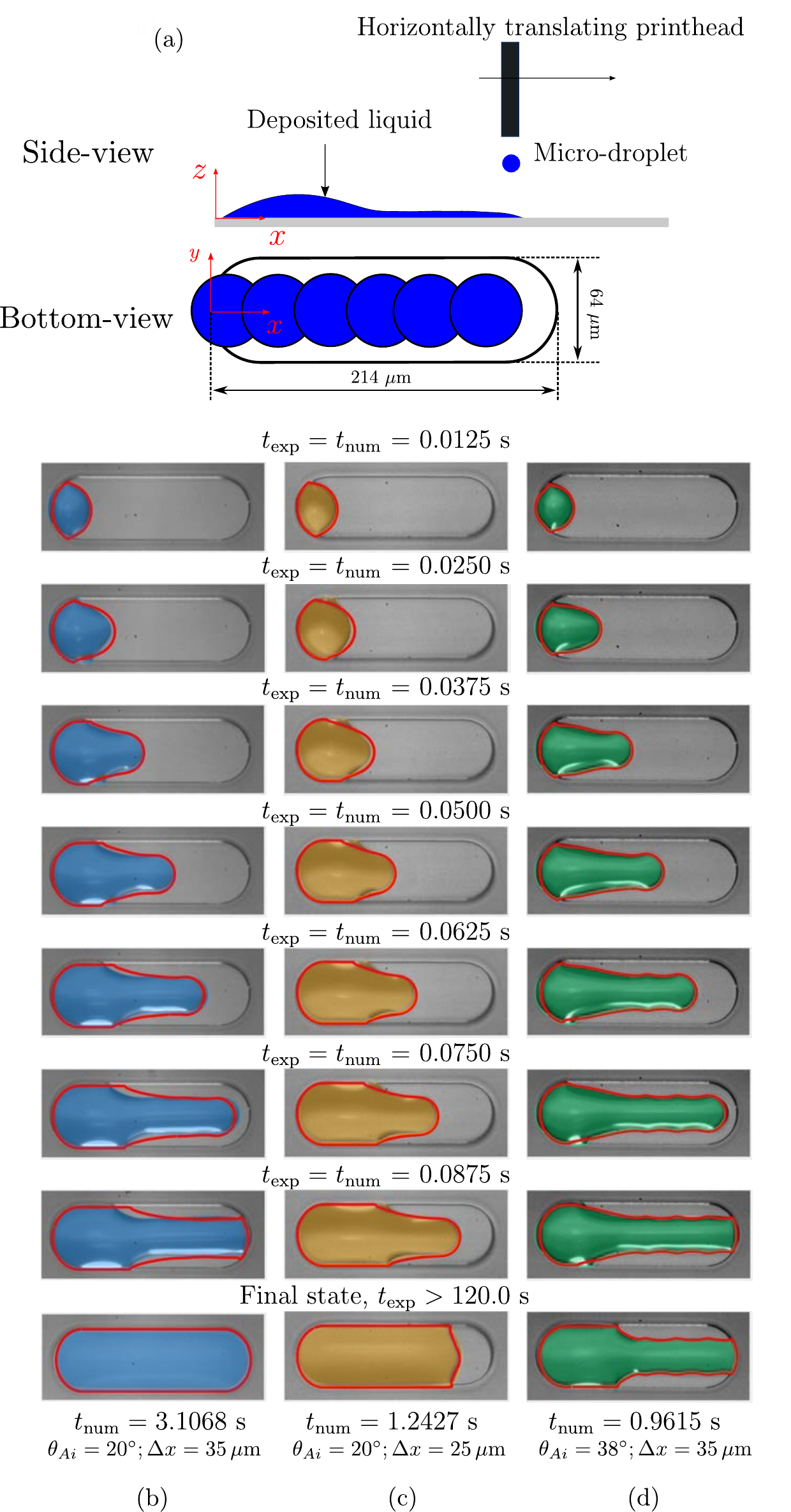}\\
\caption{(a) Schematic diagram of the experimental setup. (b,c,d) Influence of the inner advancing contact angle $\theta_{Ai}$ and inter-drop distance $\Delta x$ on the spreading of a liquid line deposited inside a pixel. Three sequences of bottom-view snapshots track the deposition of seven droplets for (b) $\theta_{Ai} = 20^\circ, \Delta x = 35\; \mu$m, (c) $\theta_{Ai} = 20^\circ, \Delta x = 25\; \mu$m and (d) $\theta_{Ai} = 38^\circ, \Delta x = 35\; \mu$m. In each sequence, the last image shows the final equilibrium configuration, taken 120.0 s after the deposition of the last droplet. The red outlines overlaid on the experimental images indicate the evolution of the liquid footprint computed numerically for the same parameters.}
\label{fig:fig1}
\end{figure}

\emph{Experimental methods}: A schematic diagram of the experimental setup is shown in Fig. 1a. The details of the experimental setup have been described elsewhere \cite{kant,kant2} and thus we will only give a brief description of the points most pertinent to the present study.
In a typical experiment equidistant partially overlapping droplets (each of volume $V$ = 7.6 $\pm$ 0.4 pL) were deposited onto a substrate using a horizontally translating drop-on-demand inkjet printhead (SX3, Fujifilm Dimatix) at a constant frequency of $f=80$~Hz.
Drops produced from the printhead had an in-flight radius $R_f = \sqrt[3]{V /(4/3\pi)} = 12.2$ $\mu$m and coalesced with the pre-existing fluid layer upon impact.
The printhead was moved horizontally at different speeds $v$ using a linear motion stage (ANT95L, Aerotech) to vary the inter-drop distance ($\Delta x = v/f$) between neighbouring droplets.
The fluid was a Cambridge Display Technology (CDT) proprietary solution used in POLED printing with dynamic viscosity $\eta$ = 6.25 $\times$ 10$^{-3}$ Pa s, density $\rho$ = 1.066 $\times$ 10$^3$ kg m$^{-3}$ and surface tension $\sigma$ = 44 $\times$ 10$^{-3}$ N m$^{-1}$.
The Weber number of the droplet impact was low, $We = 2 \rho V_i^2 R_f/\sigma = 2.4$, where $V_i=2.2$~m~s$^{-1}$ is the vertical velocity of the droplet \cite{alice}, and accordingly no splashing was observed at the time of impact \cite{visser}.
We report observations based on the bottom views recorded at 500 frames per second.

Experiments were performed on sample display backplanes manufactured in the clean room facility of CDT, each having a pixel density of 200 pixels per inch -- see Kant et al.\cite{kant2} for details of the fabrication process and techniques used for the characterisation of the topographical and wettability patterns. Each pixel on the display consisted of a recessed stadium-shaped well bound by $1.2\,\mu$m high sloping side walls (i.e. $10\%$ of the in-flight droplet radius), with
different wetting properties inside ($4^\circ \le \theta_{Ai} \le 40^\circ$, $\theta_{Ri} =0^\circ$), on the narrow sloping walls ($\theta_{Aw} =47^\circ$, $\theta_{Rw}=0^\circ$)  and outside ($\theta_{Ao}=65 \pm 2^\circ$, $\theta_{Ro}=40 \pm 2^\circ$) the pixel boundary; here $\theta_{Ax}$ and $\theta_{Rx}$ refer to the advancing and receding contact angles, respectively; see ESI. We varied the inner advancing contact angle by letting the substrate age in the laboratory; this aging did not affect the contact angles on the walls and outer surfaces which remained approximately constant. The high values of the advancing and receding contact angles on the banks surrounding the pixels ensured that any droplet deposited partially outside the boundary retreated into the pixel so that the deposited volume was contained inside the pixel \cite{kant2}.
%The `\emph{in-situ} measurement of the wetting properties of the sloped side-walls was not feasible experimentally due to the narrow width of the side wall ($\leq 1~\mu$m).
%Therefore, separate substrates were fabricated using the proprietary photo resist that formed the side wall, and both advancing and receding contact angles were measured on flat regions of these substrates to be $\theta_{Aw} (\theta_{Rw}) =47^\circ (0^\circ)$. 
%Further, we noted that the photolithographic process employed to fabricate substrates induced inhomogeneities in the wetting properties of the side-walls between pixels across the display backplane. We will address their effect on the equilibrium state of the deposited liquid in Fig. \ref{fig:fig3}.
%This wettability variation is characterised by advancing and receding contact angles measured within the pixel boundary ($4^\circ \le \theta_{Ai} \le 40^\circ$, $\theta_{Ri} =0^\circ$), on the elevated surfaces (banks) outside the pixel boundary ($\theta_{Ao}=65 \pm 2^\circ$, $\theta_{Ro}=40 \pm 2^\circ$), and on the sloping side-walls ($\theta_{Aw}=47 \pm 2^\circ$, $\theta_{Rw}=0^\circ$). 

\begin{figure}[htb]
\centering
\includegraphics[clip, trim=0.1cm 0cm 0.1cm 0cm, width=0.45\textwidth]{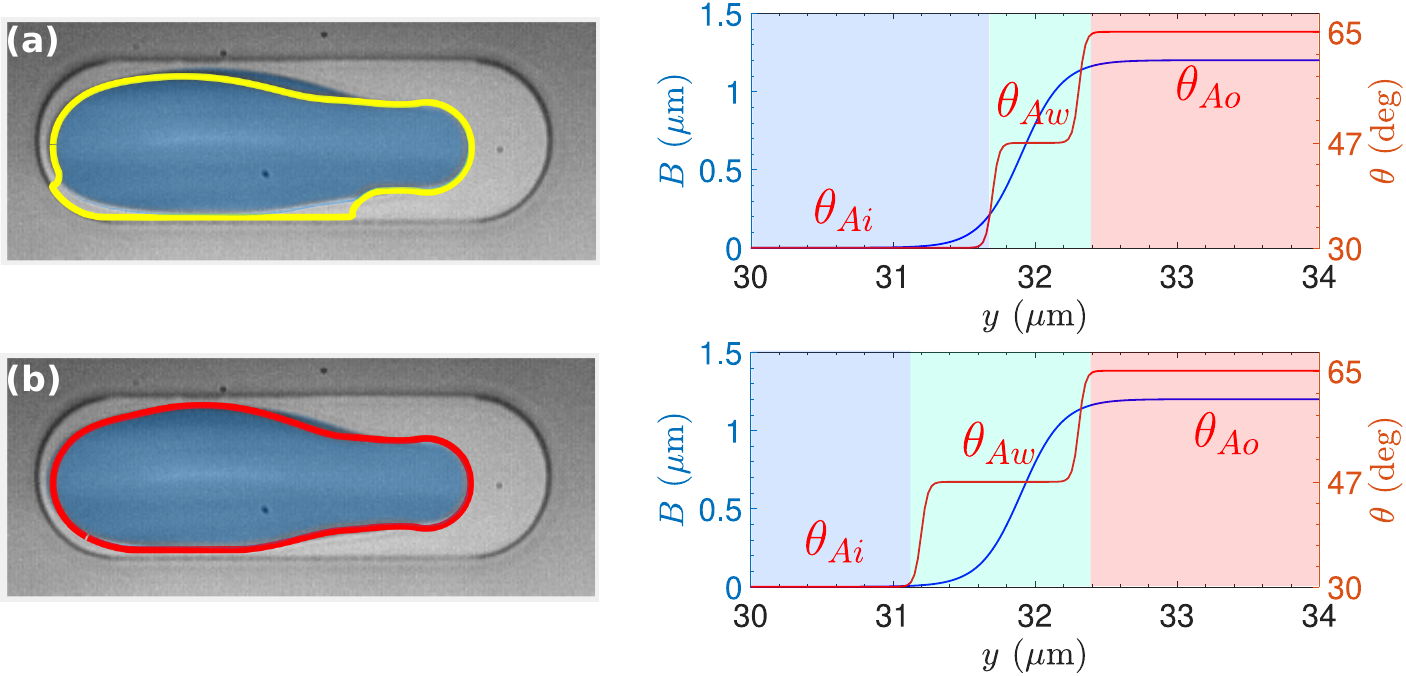}
\caption{Example of an anomalous experimental equilibrium footprint obtained following the deposition of five droplets ($\Delta x = 35\; \mu$m) with their centroids displaced from the horizontal centreline of the pixel by 5\% of the pixel width ($\theta_{Ai} = 30^\circ$). The contact line does not extend to the side walls of the pixel, suggesting that the inner-surface of the pixel is bound by a non-wetting region. This experimental footprint is compared with numerical simulations performed for (a) the wetting profile used throughout this paper (yellow contour); (b) a profile where the region with $\theta_{Aw}=47^\circ$ has been extended into the flat recessed portion of the pixel (red contour). In the wettability profiles, $B$ denotes the local height of the pixel, with $B=0$ inside the pixel, and $\theta_A$ is the advancing contact angle.}
\label{fig:fig3}
\end{figure}

\emph{Numerical model}: The details of the model development have also been presented previously\cite{alice,kant,kant2}. Briefly, we model the deposited liquid mass as a thin film of fixed volume that evolves through (quasi-static) capillary effects.  
A kinematic boundary condition advances the contact line according to
a Cox-Voinov spreading law\cite{cox,voinov}, which we parameterise
using the experimental contact-angle measurements to give a
wetting profile for  the pixel, see Fig. \ref{fig:fig3}.
The only initial conditions required for the model are the total volume and the footprint shape of the wetted region.
Upon deposition of a new droplet, 
the liquid footprint is taken as the union of the regions wetted by the existing liquid morphology and by the newly deposited droplet, and the total volume is incremented accordingly.
The fluid pressure and height profile consistent with the updated footprint and fluid volume are then adjusted to ensure that the fluid is in static equilibrium. This model was implemented numerically using the finite element library oomph-lib \cite{oomph}; see source code \cite{code} and ESI for model details.

The three sequences of images in Fig. \ref{fig:fig1} illustrate the
evolution of a liquid line formed from the deposition inside a pixel
of seven equidistant droplets, thus a fixed volume of fluid, for
different values of pixel wettability ($\theta_{Ai}$) and inter-drop
spacing ($\Delta x$). The centre of the initial droplet was always located at a
distance of $\delta x = 14 \pm 3 \mu$m from the edge of the pixel. In all three cases, the droplet deposition leads to the formation of a rivulet that fills the pixel at the upstream end and narrows as further droplets are deposited downstream. The three liquid lines evolve towards distinct equilibrium configurations (last image in each sequence), with only Fig.~\ref{fig:fig1}b resulting in a filled pixel.  
 %In all cases, the spreading of liquid within the pixel is enhanced
 %by the presence of the sloping side-walls, which locally increase
 %the dynamic contact angle, so that a wetting front propagates along
 %the pixel boundary\cite{kant2}. In conjunction with the confinement
 %imposed by the wettability patterning, this effect enables the fluid
 %to fill the pixel in Fig. \ref{fig:fig1}b, where the wetting
%conditions are favourable ($\theta_{Ai} = 20^\circ$).
In Fig. \ref{fig:fig1}d the pixel does not fill because of the increased advancing contact angle on the inner pixel surface ($\theta_{Ai} = 38^\circ$).
 In Fig. \ref{fig:fig1}b,d, the value of $\Delta x$ has been chosen so that the liquid line extends over the whole length of the pixel ($\Delta x = $35 $\mu$m) in contrast with Fig. \ref{fig:fig1}c where $\Delta x = $25 $\mu$m. Although in Fig. \ref{fig:fig1}c the contact angle is low ($\theta_{Ai} = 20^\circ$), the pixel does not fill because the line of deposited fluid does not extend to the end of the pixel (and the wetting front propagating along the pixel boundary comes to rest when it reaches the end of the liquid line).

The red contours overlaid on the experimental snapshots in
Fig. \ref{fig:fig1} represent the  footprints of the deposited liquid
computed numerically for the same parameter values as in the
experimental sequences. The model captures the main features of the
evolution of the liquid lines for different values of $\theta_{Ai}$
and $\Delta x$. At early times ($t \le 0.0875$ s), the model
predictions are almost indistinguishable from the experiment. Small
discrepancies in the position of the wetting fronts along the
side-walls occur only in Fig. \ref{fig:fig1}b, which we attribute to
experimental variability in the wetting properties of the side-walls
(see Fig. \ref{fig:fig3} for further discussion). At later times, the
liquid spreads faster in the computations because the model does not
include any viscous effects which begin to restrict the flow in
the experiments. Despite the difference in the spreading time scales, the model correctly predicts the evolution and the equilibrium states of the three liquid lines in Fig. \ref{fig:fig1}.

\begin{figure*}[htb]
\centering
\includegraphics[width=0.8\textwidth]{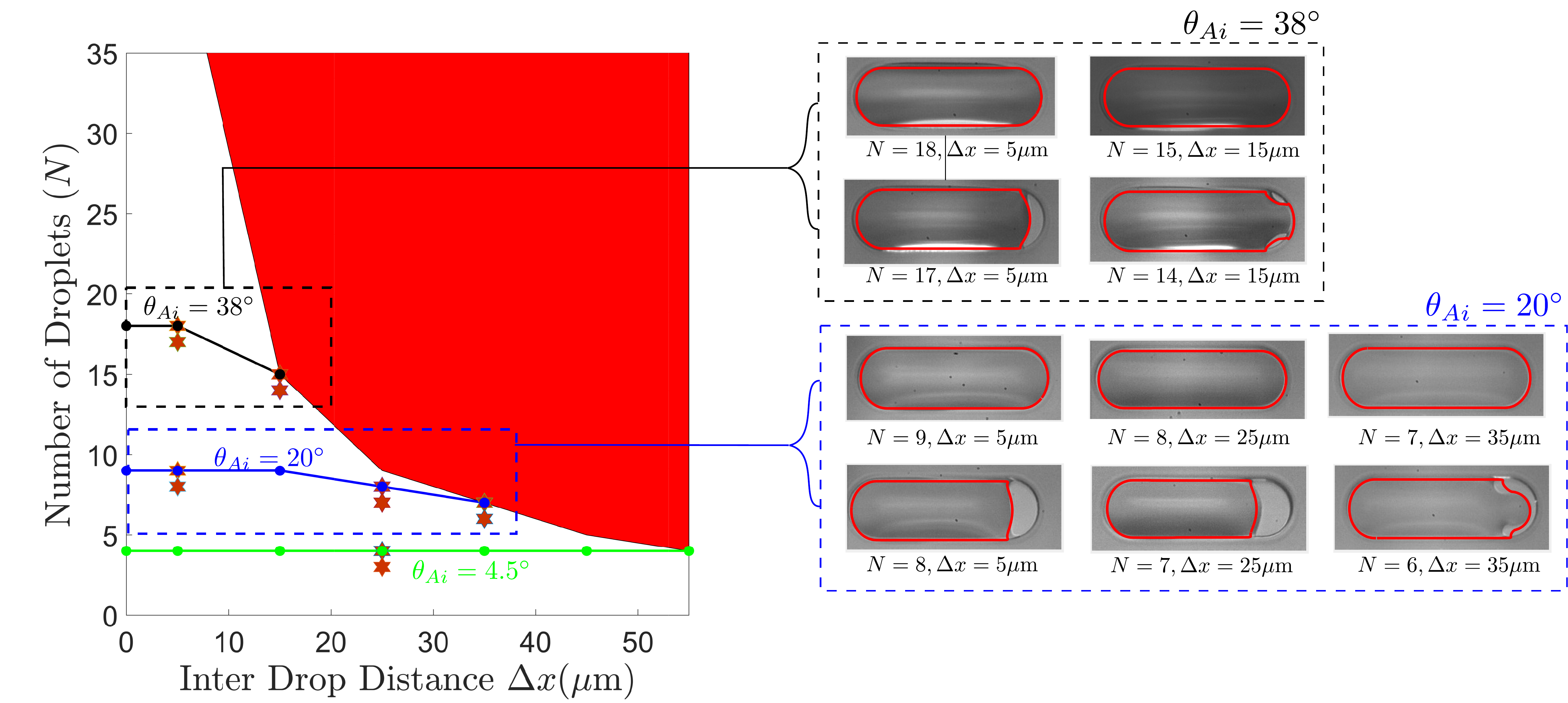}\\
\caption{Phase diagram summarizing the influence of the inner advancing contact angle $\theta_{Ai}$ and the printing protocol ($N, \Delta x$) on the filling of a stadium-shaped pixel. The boundary of the red-shaded region represents the maximum number of droplets which can be deposited at least partially inside the pixel for a chosen value of $\Delta x$. The filled circles, linearly interpolated by solid lines, correspond to the threshold operating conditions that separate filled and partially filled pixels with $\theta_{Ai} = 4.5^\circ$ (green), $20^\circ$ (blue) and $38^\circ$ (black), respectively. Each data point indicates three concurrent experiments, and the accuracy of each data point was confirmed by performing repeated experiments for one and two fewer droplets as well as for one more droplet, which consistently led to partially filled and filled pixels, respectively. Hence, safe operating conditions that result in a filled pixel are in the regions above the threshold conditions and left of the red-shaded region. The stars correspond to the operating conditions of the snapshots on the right side of the graph for $\theta_{Ai} = 20^\circ; \; 38^\circ$, and in the ESI for $\theta_{Ai} = 4.5^\circ$. The red contours are the numerically computed equilibrium footprints for the same parameter values.}
\label{fig:fig2}
\end{figure*}

The examples shown in Fig. \ref{fig:fig1} demonstrate that the filling
of pixels does not occur for all values of $\theta_{Ai}$ and $\Delta
x$. A large number of experiments ($\sim 100$) were conducted to construct a phase diagram of `filled' or `partially filled'
equilibrium states in terms of the operating conditions ($N$, $\Delta
x$) for pixels of three different wettabilities $\theta_{Ai} =
4.5^\circ,\,20^\circ$ and $38^\circ$. The resulting phase diagram is
shown in Fig. \ref{fig:fig2}. The shaded region indicates conditions
for which at least one droplet is deposited entirely outside the
pixel, while the filled circles interpolated by solid lines represent the minimum
number of droplets required to fill a pixel as a function of
inter-drop distance, for different values of $\theta_{Ai}$.
Thus, the areas above each threshold boundary that are bordered by the
red-shaded region on the right hand-side correspond to the `safe'
operating conditions that systematically lead to filled pixels. 
Note that the boundaries between different regions in the phase diagram were determined experimentally.

For wetting pixels ($\theta_{Ai}=4.5^\circ$), `safe' operating conditions
are for $N \ge 4$ (green line) because the low contact angle enables
the line to spread to the end of the pixel regardless of droplet
positioning. This means that the filling of a wetting pixel only
depends on the total volume deposited; see Figs. S4 and S5.
The number of droplets required to fill the pixel decreases with decreasing contact
angle until for strongly wetting substrates ($\theta_{A} < 1^\circ$)
a single droplet can fill the pixel provided that the volumetric evaporation rate is not too high. 

For partially wetting substrates ($20^\circ \leq \theta \leq 38^\circ$), the filling of a pixel for small values of $\Delta x$ also requires the volume deposited to exceed a threshold value. However, the number of droplets required to fill the pixel decreases as the inter-drop distance increases (blue line: $\theta_{Ai}=20^\circ$; black line: $\theta_{Ai}=38^\circ$) because the filling of the pixel relies on the liquid line reaching the downstream end of the pixel and this can be achieved for a reduced total deposited volume by increasing $\Delta x$. This result is consistent with Fig. \ref{fig:fig1}b, where at $\Delta x=35\;\mu$m, the liquid line extends over the length of the pixel and the side walls enhance spreading sufficiently to fill the pixel. Overall, the area of the parameter plane that leads to filled equilibrium states shrinks with increasing $\theta_{Ai}$. However, the minimum number of droplets required to fill the pixel can be reduced by using droplets of higher volume, thus expanding the safe range of operating conditions; see Fig. S6. 

Using the numerical model, we were able to reproduce the threshold operating conditions that separate filled and partially filled pixel states; see Fig. \ref{fig:fig2}. 
%The symbols in Fig. \ref{fig:fig2} denote comparisons between experiments and numerical computations on the blue and black lines in the phase diagram (filled pixels) and one drop below these lines (partially filled pixels). 
The numerical model systematically reproduces the details of the
equilibrium configurations observed in the experiments, and hence
predicts the boundary between filled and under-filled pixels.
This agreement demonstrates the utility of our computationally inexpensive model,
promoting it as a feasible design tool for inkjet-printing-based
manufacturing.

In a small subset of experimental pixels, anomalous equilibrium
configurations were obtained, which the model used in
Figs.~\ref{fig:fig1} and \ref{fig:fig2} failed to capture. An example
is shown in Fig. \ref{fig:fig3}, where the off-centred deposition of
five droplets yields an equilibrium liquid line that does not extend
to the side walls. The footprint calculated numerically with the
wetting profile used so far, differs from the experiment in that one
side of the footprint has spread along the bounding wall, as indicated
by the presence of a wetting front (Fig. \ref{fig:fig3}a). We find
that a change in the imposed wetting profile can reproduce the
experimental equilibrium state numerically (Fig. \ref{fig:fig3}b): the
region where  $\theta_{Aw}(\theta_{Rw}) = 47^\circ(0^\circ)$ is
extended by 0.5 $\mu$m (1.5\% of the pixel width) into the inner pixel
region. This result suggests that the numerical model can be used to diagnose small variations in the wetting properties of the side-walls across pixels, which result from the fabrication process. This known variability in the wetting profile can in turn be accounted for when numerically predicting safe operating conditions that lead to filled pixels.

%Experiments revealed that the equilibrium morphology of a deposited liquid line depends sensitively on the wettability profile near the sloping side-walls, which controls the formation of wetting fronts along the side-walls. the experimental footprints were obtained  that
%, whereas, in Fig. \ref{fig:fig3}b, the corner formed between the flat inner pixel surface and the sloping side-wall was modelled as part of the inner-pixel surface with contact angles $\theta_{Ai}$ and $\theta_{Ri}$, thus enabling the development of a wetting front to spread along the pixel boundary.
%Aforementioned, the variability in the wettability profile on these small length scales stems from the photo-lithographic process employed to fabricate display backplanes and hence is unavoidable. Thus, our numerical model can serve as an effective design tool to determine robust operating conditions that guarantee pixel filling in the presence of known variabilities in the wetting profile near the pixel side-walls.

%deposition of five droplets with $theta_i=30^\circ$ and $\Delta x=35\;\mu$m. In this case. The use of the wetting profile.

\begin{figure}
\centering
\includegraphics[width=0.45\textwidth]{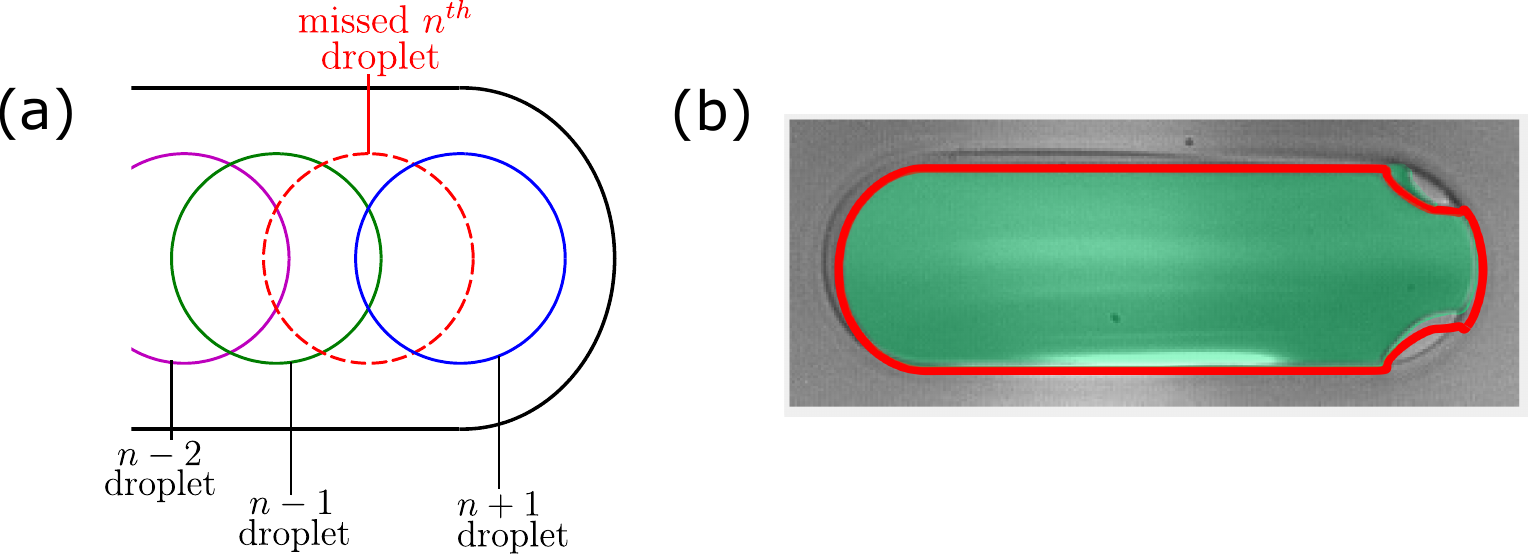}\\
\caption{(a) Schematic diagram illustrating the numerical scheme used to simulate a common printhead dysfunction (missed droplet). (b) Comparison between experimental and numerical footprints of an equilibrium configuration following a deposition sequence where the 13$^{\mathrm{th}}$ droplet out of 14 was missed out: $\theta_{Ai} = 38^\circ$, $\Delta x$ = 15 $\mu$m.}
\label{fig:fig4}
\end{figure}

Finally, we used the numerical model to capture the effect of a printhead dysfunction commonly encountered during operations on an industrial scale, namely the misfiring of a droplet.
As printing defects can lead to partially filled pixels\cite{halls2005ink} which reduce the overall resolution and performance of a display device, it is necessary to account for such dysfunctions at the design stage. In Fig. \ref{fig:fig4}, a numerical scheme is depicted which misses out the printing of the penultimate droplet, in order to account for the misfiring of a droplet during the deposition sequence. Experimental and numerical equilibrium configurations obtained following this protocol are in excellent agreement. This test suggests that the numerical model may be used to identify a subset of the safe operating conditions reported in Fig. \ref{fig:fig2} where filling is ensured despite the occasional misfiring of the printhead.
%\textbf{Alice highlighted that we may want to add an extra comparison here, but in my view it is alright at moment. However, if we decide to add more material here, I can add it later.}

To summarize, we have discussed various aspects pertaining to the printing of short liquid lines to robustly fill pixels in the context of POLED display fabrication.
We have shown that in order for sequentially deposited droplets that coalesce upon impact to form a thin film that covers the entire pixel surface, the wettability of the pixel surface must be considered in conjunction with the deposition parameters.
Furthermore, we have demonstrated that a simple model that relies on capillary spreading according to a Cox-Voinov spreading law can be used as a computationally inexpensive design tool to determine efficient printing strategies in the fabrication of high performance POLED and also LCD (Liquid Crystal Display)\cite{LCD} devices.

%\nocite{*}
\bibliography{bib}% Produces the bibliography via BibTeX.

\providecommand{\noopsort}[1]{}\providecommand{\singleletter}[1]{#1}%
\begin{thebibliography}{33}%
\makeatletter
\providecommand \@ifxundefined [1]{%
 \@ifx{#1\undefined}
}%
\providecommand \@ifnum [1]{%
 \ifnum #1\expandafter \@firstoftwo
 \else \expandafter \@secondoftwo
 \fi
}%
\providecommand \@ifx [1]{%
 \ifx #1\expandafter \@firstoftwo
 \else \expandafter \@secondoftwo
 \fi
}%
\providecommand \natexlab [1]{#1}%
\providecommand \enquote  [1]{``#1''}%
\providecommand \bibnamefont  [1]{#1}%
\providecommand \bibfnamefont [1]{#1}%
\providecommand \citenamefont [1]{#1}%
\providecommand \href@noop [0]{\@secondoftwo}%
\providecommand \href [0]{\begingroup \@sanitize@url \@href}%
\providecommand \@href[1]{\@@startlink{#1}\@@href}%
\providecommand \@@href[1]{\endgroup#1\@@endlink}%
\providecommand \@sanitize@url [0]{\catcode `\\12\catcode `\$12\catcode
  `\&12\catcode `\#12\catcode `\^12\catcode `\_12\catcode `\%12\relax}%
\providecommand \@@startlink[1]{}%
\providecommand \@@endlink[0]{}%
\providecommand \url  [0]{\begingroup\@sanitize@url \@url }%
\providecommand \@url [1]{\endgroup\@href {#1}{\urlprefix }}%
\providecommand \urlprefix  [0]{URL }%
\providecommand \Eprint [0]{\href }%
\providecommand \doibase [0]{http://dx.doi.org/}%
\providecommand \selectlanguage [0]{\@gobble}%
\providecommand \bibinfo  [0]{\@secondoftwo}%
\providecommand \bibfield  [0]{\@secondoftwo}%
\providecommand \translation [1]{[#1]}%
\providecommand \BibitemOpen [0]{}%
\providecommand \bibitemStop [0]{}%
\providecommand \bibitemNoStop [0]{.\EOS\space}%
\providecommand \EOS [0]{\spacefactor3000\relax}%
\providecommand \BibitemShut  [1]{\csname bibitem#1\endcsname}%
\let\auto@bib@innerbib\@empty
%</preamble>
\bibitem [{\citenamefont {Calvert}(2001)}]{calvert2001inkjet}%
  \BibitemOpen
  \bibfield  {author} {\bibinfo {author} {\bibfnamefont {P.}~\bibnamefont
  {Calvert}},\ }\href@noop {} {\bibfield  {journal} {\bibinfo  {journal} {Chem.
  Mater.}\ }\textbf {\bibinfo {volume} {13}},\ \bibinfo {pages} {3299}
  (\bibinfo {year} {2001})}\BibitemShut {NoStop}%
\bibitem [{\citenamefont {Singh}\ \emph {et~al.}(2010)\citenamefont {Singh},
  \citenamefont {Haverinen}, \citenamefont {Dhagat},\ and\ \citenamefont
  {Jabbour}}]{singh2010inkjet}%
  \BibitemOpen
  \bibfield  {author} {\bibinfo {author} {\bibfnamefont {M.}~\bibnamefont
  {Singh}}, \bibinfo {author} {\bibfnamefont {H.~M.}\ \bibnamefont
  {Haverinen}}, \bibinfo {author} {\bibfnamefont {P.}~\bibnamefont {Dhagat}}, \
  and\ \bibinfo {author} {\bibfnamefont {G.~E.}\ \bibnamefont {Jabbour}},\
  }\href@noop {} {\bibfield  {journal} {\bibinfo  {journal} {Adv. Maters.}\
  }\textbf {\bibinfo {volume} {22}},\ \bibinfo {pages} {673} (\bibinfo {year}
  {2010})}\BibitemShut {NoStop}%
\bibitem [{\citenamefont {Sirringhaus}\ \emph {et~al.}(2000)\citenamefont
  {Sirringhaus}, \citenamefont {Kawase}, \citenamefont {Friend}, \citenamefont
  {Shimoda}, \citenamefont {Inbasekaran}, \citenamefont {Wu},\ and\
  \citenamefont {Woo}}]{sirringhaus2000high}%
  \BibitemOpen
  \bibfield  {author} {\bibinfo {author} {\bibfnamefont {H.}~\bibnamefont
  {Sirringhaus}}, \bibinfo {author} {\bibfnamefont {T.}~\bibnamefont {Kawase}},
  \bibinfo {author} {\bibfnamefont {R.}~\bibnamefont {Friend}}, \bibinfo
  {author} {\bibfnamefont {T.}~\bibnamefont {Shimoda}}, \bibinfo {author}
  {\bibfnamefont {M.}~\bibnamefont {Inbasekaran}}, \bibinfo {author}
  {\bibfnamefont {W.}~\bibnamefont {Wu}}, \ and\ \bibinfo {author}
  {\bibfnamefont {E.}~\bibnamefont {Woo}},\ }\href@noop {} {\bibfield
  {journal} {\bibinfo  {journal} {Science}\ }\textbf {\bibinfo {volume}
  {290}},\ \bibinfo {pages} {2123} (\bibinfo {year} {2000})}\BibitemShut
  {NoStop}%
\bibitem [{\citenamefont {Minemawari}\ \emph {et~al.}(2011)\citenamefont
  {Minemawari}, \citenamefont {Yamada}, \citenamefont {Matsui}, \citenamefont
  {Tsutsumi}, \citenamefont {Haas}, \citenamefont {Chiba}, \citenamefont
  {Kumai},\ and\ \citenamefont {Hasegawa}}]{minemawari2011inkjet}%
  \BibitemOpen
  \bibfield  {author} {\bibinfo {author} {\bibfnamefont {H.}~\bibnamefont
  {Minemawari}}, \bibinfo {author} {\bibfnamefont {T.}~\bibnamefont {Yamada}},
  \bibinfo {author} {\bibfnamefont {H.}~\bibnamefont {Matsui}}, \bibinfo
  {author} {\bibfnamefont {J.}~\bibnamefont {Tsutsumi}}, \bibinfo {author}
  {\bibfnamefont {S.}~\bibnamefont {Haas}}, \bibinfo {author} {\bibfnamefont
  {R.}~\bibnamefont {Chiba}}, \bibinfo {author} {\bibfnamefont
  {R.}~\bibnamefont {Kumai}}, \ and\ \bibinfo {author} {\bibfnamefont
  {T.}~\bibnamefont {Hasegawa}},\ }\href@noop {} {\bibfield  {journal}
  {\bibinfo  {journal} {Nature}\ }\textbf {\bibinfo {volume} {475}},\ \bibinfo
  {pages} {364} (\bibinfo {year} {2011})}\BibitemShut {NoStop}%
\bibitem [{\citenamefont {Ko}\ \emph {et~al.}(2007)\citenamefont {Ko},
  \citenamefont {Pan}, \citenamefont {Grigoropoulos}, \citenamefont {Luscombe},
  \citenamefont {Fr{\'e}chet},\ and\ \citenamefont {Poulikakos}}]{ko2007all}%
  \BibitemOpen
  \bibfield  {author} {\bibinfo {author} {\bibfnamefont {S.~H.}\ \bibnamefont
  {Ko}}, \bibinfo {author} {\bibfnamefont {H.}~\bibnamefont {Pan}}, \bibinfo
  {author} {\bibfnamefont {C.~P.}\ \bibnamefont {Grigoropoulos}}, \bibinfo
  {author} {\bibfnamefont {C.~K.}\ \bibnamefont {Luscombe}}, \bibinfo {author}
  {\bibfnamefont {J.~M.}\ \bibnamefont {Fr{\'e}chet}}, \ and\ \bibinfo {author}
  {\bibfnamefont {D.}~\bibnamefont {Poulikakos}},\ }\href@noop {} {\bibfield
  {journal} {\bibinfo  {journal} {Nanotechnology}\ }\textbf {\bibinfo {volume}
  {18}},\ \bibinfo {pages} {345202} (\bibinfo {year} {2007})}\BibitemShut
  {NoStop}%
\bibitem [{\citenamefont {Andersson}\ \emph {et~al.}(2007)\citenamefont
  {Andersson}, \citenamefont {Forchheimer}, \citenamefont {Tehrani},\ and\
  \citenamefont {Berggren}}]{andersson2007printable}%
  \BibitemOpen
  \bibfield  {author} {\bibinfo {author} {\bibfnamefont {P.}~\bibnamefont
  {Andersson}}, \bibinfo {author} {\bibfnamefont {R.}~\bibnamefont
  {Forchheimer}}, \bibinfo {author} {\bibfnamefont {P.}~\bibnamefont
  {Tehrani}}, \ and\ \bibinfo {author} {\bibfnamefont {M.}~\bibnamefont
  {Berggren}},\ }\href@noop {} {\bibfield  {journal} {\bibinfo  {journal} {Adv.
  Funct. Mater.}\ }\textbf {\bibinfo {volume} {17}},\ \bibinfo {pages} {3074}
  (\bibinfo {year} {2007})}\BibitemShut {NoStop}%
\bibitem [{\citenamefont {Shimoda}\ \emph {et~al.}(2003)\citenamefont
  {Shimoda}, \citenamefont {Morii}, \citenamefont {Seki},\ and\ \citenamefont
  {Kiguchi}}]{shimoda2003inkjet}%
  \BibitemOpen
  \bibfield  {author} {\bibinfo {author} {\bibfnamefont {T.}~\bibnamefont
  {Shimoda}}, \bibinfo {author} {\bibfnamefont {K.}~\bibnamefont {Morii}},
  \bibinfo {author} {\bibfnamefont {S.}~\bibnamefont {Seki}}, \ and\ \bibinfo
  {author} {\bibfnamefont {H.}~\bibnamefont {Kiguchi}},\ }\href@noop {}
  {\bibfield  {journal} {\bibinfo  {journal} {MRS Bull.}\ }\textbf {\bibinfo
  {volume} {28}},\ \bibinfo {pages} {821} (\bibinfo {year} {2003})}\BibitemShut
  {NoStop}%
\bibitem [{\citenamefont {Halls}(2005)}]{halls2005ink}%
  \BibitemOpen
  \bibfield  {author} {\bibinfo {author} {\bibfnamefont {J.}~\bibnamefont
  {Halls}},\ }in\ \href@noop {} {\emph {\bibinfo {booktitle} {Inf. Disp}}},\
  Vol.~\bibinfo {volume} {2}\ (\bibinfo {year} {2005})\BibitemShut {NoStop}%
\bibitem [{\citenamefont {Dua}\ \emph {et~al.}(2010)\citenamefont {Dua},
  \citenamefont {Surwade}, \citenamefont {Ammu}, \citenamefont {Agnihotra},
  \citenamefont {Jain}, \citenamefont {Roberts}, \citenamefont {Park},
  \citenamefont {Ruoff},\ and\ \citenamefont {Manohar}}]{dua2010all}%
  \BibitemOpen
  \bibfield  {author} {\bibinfo {author} {\bibfnamefont {V.}~\bibnamefont
  {Dua}}, \bibinfo {author} {\bibfnamefont {S.~P.}\ \bibnamefont {Surwade}},
  \bibinfo {author} {\bibfnamefont {S.}~\bibnamefont {Ammu}}, \bibinfo {author}
  {\bibfnamefont {S.~R.}\ \bibnamefont {Agnihotra}}, \bibinfo {author}
  {\bibfnamefont {S.}~\bibnamefont {Jain}}, \bibinfo {author} {\bibfnamefont
  {K.~E.}\ \bibnamefont {Roberts}}, \bibinfo {author} {\bibfnamefont
  {S.}~\bibnamefont {Park}}, \bibinfo {author} {\bibfnamefont {R.~S.}\
  \bibnamefont {Ruoff}}, \ and\ \bibinfo {author} {\bibfnamefont {S.~K.}\
  \bibnamefont {Manohar}},\ }\href@noop {} {\bibfield  {journal} {\bibinfo
  {journal} {Angew. Chem.}\ }\textbf {\bibinfo {volume} {122}},\ \bibinfo
  {pages} {2200} (\bibinfo {year} {2010})}\BibitemShut {NoStop}%
\bibitem [{\citenamefont {Lorwongtragool}\ \emph {et~al.}(2014)\citenamefont
  {Lorwongtragool}, \citenamefont {Sowade}, \citenamefont {Watthanawisuth},
  \citenamefont {Baumann},\ and\ \citenamefont
  {Kerdcharoen}}]{lorwongtragool2014novel}%
  \BibitemOpen
  \bibfield  {author} {\bibinfo {author} {\bibfnamefont {P.}~\bibnamefont
  {Lorwongtragool}}, \bibinfo {author} {\bibfnamefont {E.}~\bibnamefont
  {Sowade}}, \bibinfo {author} {\bibfnamefont {N.}~\bibnamefont
  {Watthanawisuth}}, \bibinfo {author} {\bibfnamefont {R.~R.}\ \bibnamefont
  {Baumann}}, \ and\ \bibinfo {author} {\bibfnamefont {T.}~\bibnamefont
  {Kerdcharoen}},\ }\href@noop {} {\bibfield  {journal} {\bibinfo  {journal}
  {Sensors}\ }\textbf {\bibinfo {volume} {14}},\ \bibinfo {pages} {19700}
  (\bibinfo {year} {2014})}\BibitemShut {NoStop}%
\bibitem [{\citenamefont {Abe}, \citenamefont {Suzuki},\ and\ \citenamefont
  {Citterio}(2008)}]{abe2008inkjet}%
  \BibitemOpen
  \bibfield  {author} {\bibinfo {author} {\bibfnamefont {K.}~\bibnamefont
  {Abe}}, \bibinfo {author} {\bibfnamefont {K.}~\bibnamefont {Suzuki}}, \ and\
  \bibinfo {author} {\bibfnamefont {D.}~\bibnamefont {Citterio}},\ }\href@noop
  {} {\bibfield  {journal} {\bibinfo  {journal} {Anal. Chem.}\ }\textbf
  {\bibinfo {volume} {80}},\ \bibinfo {pages} {6928} (\bibinfo {year}
  {2008})}\BibitemShut {NoStop}%
\bibitem [{\citenamefont {Hoth}\ \emph {et~al.}(2008)\citenamefont {Hoth},
  \citenamefont {Schilinsky}, \citenamefont {Choulis},\ and\ \citenamefont
  {Brabec}}]{hoth2008printing}%
  \BibitemOpen
  \bibfield  {author} {\bibinfo {author} {\bibfnamefont {C.~N.}\ \bibnamefont
  {Hoth}}, \bibinfo {author} {\bibfnamefont {P.}~\bibnamefont {Schilinsky}},
  \bibinfo {author} {\bibfnamefont {S.~A.}\ \bibnamefont {Choulis}}, \ and\
  \bibinfo {author} {\bibfnamefont {C.~J.}\ \bibnamefont {Brabec}},\
  }\href@noop {} {\bibfield  {journal} {\bibinfo  {journal} {Nano Lett.}\
  }\textbf {\bibinfo {volume} {8}},\ \bibinfo {pages} {2806} (\bibinfo {year}
  {2008})}\BibitemShut {NoStop}%
\bibitem [{\citenamefont {Jung}\ \emph {et~al.}(2014)\citenamefont {Jung},
  \citenamefont {Sou}, \citenamefont {Banger}, \citenamefont {Ko},
  \citenamefont {Chow}, \citenamefont {McNeill},\ and\ \citenamefont
  {Sirringhaus}}]{jung2014all}%
  \BibitemOpen
  \bibfield  {author} {\bibinfo {author} {\bibfnamefont {S.}~\bibnamefont
  {Jung}}, \bibinfo {author} {\bibfnamefont {A.}~\bibnamefont {Sou}}, \bibinfo
  {author} {\bibfnamefont {K.}~\bibnamefont {Banger}}, \bibinfo {author}
  {\bibfnamefont {D.-H.}\ \bibnamefont {Ko}}, \bibinfo {author} {\bibfnamefont
  {P.~C.}\ \bibnamefont {Chow}}, \bibinfo {author} {\bibfnamefont {C.~R.}\
  \bibnamefont {McNeill}}, \ and\ \bibinfo {author} {\bibfnamefont
  {H.}~\bibnamefont {Sirringhaus}},\ }\href@noop {} {\bibfield  {journal}
  {\bibinfo  {journal} {Adv. Energy Mater.}\ }\textbf {\bibinfo {volume} {4}},\
  \bibinfo {pages} {1400432} (\bibinfo {year} {2014})}\BibitemShut {NoStop}%
\bibitem [{\citenamefont {Stringer}\ and\ \citenamefont
  {Derby}(2010)}]{stringer2010formation}%
  \BibitemOpen
  \bibfield  {author} {\bibinfo {author} {\bibfnamefont {J.}~\bibnamefont
  {Stringer}}\ and\ \bibinfo {author} {\bibfnamefont {B.}~\bibnamefont
  {Derby}},\ }\href@noop {} {\bibfield  {journal} {\bibinfo  {journal}
  {Langmuir}\ }\textbf {\bibinfo {volume} {26}},\ \bibinfo {pages} {10365}
  (\bibinfo {year} {2010})}\BibitemShut {NoStop}%
\bibitem [{\citenamefont {Duineveld}(2003)}]{duineveld2003stability}%
  \BibitemOpen
  \bibfield  {author} {\bibinfo {author} {\bibfnamefont {P.~C.}\ \bibnamefont
  {Duineveld}},\ }\href@noop {} {\bibfield  {journal} {\bibinfo  {journal} {J.
  Fluid Mech.}\ }\textbf {\bibinfo {volume} {477}},\ \bibinfo {pages} {175}
  (\bibinfo {year} {2003})}\BibitemShut {NoStop}%
\bibitem [{\citenamefont {Dalili}\ \emph {et~al.}(2014)\citenamefont {Dalili},
  \citenamefont {Chandra}, \citenamefont {Mostaghimi}, \citenamefont {Fan},\
  and\ \citenamefont {Simmer}}]{dalili2014formation}%
  \BibitemOpen
  \bibfield  {author} {\bibinfo {author} {\bibfnamefont {A.}~\bibnamefont
  {Dalili}}, \bibinfo {author} {\bibfnamefont {S.}~\bibnamefont {Chandra}},
  \bibinfo {author} {\bibfnamefont {J.}~\bibnamefont {Mostaghimi}}, \bibinfo
  {author} {\bibfnamefont {H.~C.}\ \bibnamefont {Fan}}, \ and\ \bibinfo
  {author} {\bibfnamefont {J.~C.}\ \bibnamefont {Simmer}},\ }\href@noop {}
  {\bibfield  {journal} {\bibinfo  {journal} {J. Colloid Interf. Sci.}\
  }\textbf {\bibinfo {volume} {418}},\ \bibinfo {pages} {292} (\bibinfo {year}
  {2014})}\BibitemShut {NoStop}%
\bibitem [{\citenamefont {Sowade}\ \emph {et~al.}(2016)\citenamefont {Sowade},
  \citenamefont {Mitra}, \citenamefont {Ramon}, \citenamefont
  {Martinez-Domingo}, \citenamefont {Villani}, \citenamefont {Loffredo},
  \citenamefont {Gomes},\ and\ \citenamefont {Baumann}}]{sowade2016up}%
  \BibitemOpen
  \bibfield  {author} {\bibinfo {author} {\bibfnamefont {E.}~\bibnamefont
  {Sowade}}, \bibinfo {author} {\bibfnamefont {K.~Y.}\ \bibnamefont {Mitra}},
  \bibinfo {author} {\bibfnamefont {E.}~\bibnamefont {Ramon}}, \bibinfo
  {author} {\bibfnamefont {C.}~\bibnamefont {Martinez-Domingo}}, \bibinfo
  {author} {\bibfnamefont {F.}~\bibnamefont {Villani}}, \bibinfo {author}
  {\bibfnamefont {F.}~\bibnamefont {Loffredo}}, \bibinfo {author}
  {\bibfnamefont {H.~L.}\ \bibnamefont {Gomes}}, \ and\ \bibinfo {author}
  {\bibfnamefont {R.~R.}\ \bibnamefont {Baumann}},\ }\href@noop {} {\bibfield
  {journal} {\bibinfo  {journal} {Org. Electron.}\ }\textbf {\bibinfo {volume}
  {30}},\ \bibinfo {pages} {237} (\bibinfo {year} {2016})}\BibitemShut
  {NoStop}%
\bibitem [{\citenamefont {Noh}\ \emph {et~al.}(2015)\citenamefont {Noh},
  \citenamefont {Jung}, \citenamefont {Jung}, \citenamefont {Yeom},
  \citenamefont {Pyo},\ and\ \citenamefont {Cho}}]{noh2015key}%
  \BibitemOpen
  \bibfield  {author} {\bibinfo {author} {\bibfnamefont {J.}~\bibnamefont
  {Noh}}, \bibinfo {author} {\bibfnamefont {M.}~\bibnamefont {Jung}}, \bibinfo
  {author} {\bibfnamefont {Y.}~\bibnamefont {Jung}}, \bibinfo {author}
  {\bibfnamefont {C.}~\bibnamefont {Yeom}}, \bibinfo {author} {\bibfnamefont
  {M.}~\bibnamefont {Pyo}}, \ and\ \bibinfo {author} {\bibfnamefont
  {G.}~\bibnamefont {Cho}},\ }\href@noop {} {\bibfield  {journal} {\bibinfo
  {journal} {Proc. IEEE}\ }\textbf {\bibinfo {volume} {103}},\ \bibinfo {pages}
  {554} (\bibinfo {year} {2015})}\BibitemShut {NoStop}%
\bibitem [{\citenamefont {Soltman}\ and\ \citenamefont
  {Subramanian}(2008)}]{soltman2008inkjet}%
  \BibitemOpen
  \bibfield  {author} {\bibinfo {author} {\bibfnamefont {D.}~\bibnamefont
  {Soltman}}\ and\ \bibinfo {author} {\bibfnamefont {V.}~\bibnamefont
  {Subramanian}},\ }\href@noop {} {\bibfield  {journal} {\bibinfo  {journal}
  {Langmuir}\ }\textbf {\bibinfo {volume} {24}},\ \bibinfo {pages} {2224}
  (\bibinfo {year} {2008})}\BibitemShut {NoStop}%
\bibitem [{\citenamefont {Thompson}\ \emph {et~al.}(2014)\citenamefont
  {Thompson}, \citenamefont {Tipton}, \citenamefont {Juel}, \citenamefont
  {Hazel},\ and\ \citenamefont {Dowling}}]{alice}%
  \BibitemOpen
  \bibfield  {author} {\bibinfo {author} {\bibfnamefont {A.~B.}\ \bibnamefont
  {Thompson}}, \bibinfo {author} {\bibfnamefont {C.~R.}\ \bibnamefont
  {Tipton}}, \bibinfo {author} {\bibfnamefont {A.}~\bibnamefont {Juel}},
  \bibinfo {author} {\bibfnamefont {A.~L.}\ \bibnamefont {Hazel}}, \ and\
  \bibinfo {author} {\bibfnamefont {M.}~\bibnamefont {Dowling}},\ }\href@noop
  {} {\bibfield  {journal} {\bibinfo  {journal} {J. Fluid Mech.}\ }\textbf
  {\bibinfo {volume} {761}},\ \bibinfo {pages} {261} (\bibinfo {year}
  {2014})}\BibitemShut {NoStop}%
\bibitem [{\citenamefont {Wang}\ \emph {et~al.}(2004)\citenamefont {Wang},
  \citenamefont {Zheng}, \citenamefont {Li}, \citenamefont {Huck},\ and\
  \citenamefont {Sirringhaus}}]{wang2004dewetting}%
  \BibitemOpen
  \bibfield  {author} {\bibinfo {author} {\bibfnamefont {J.}~\bibnamefont
  {Wang}}, \bibinfo {author} {\bibfnamefont {Z.}~\bibnamefont {Zheng}},
  \bibinfo {author} {\bibfnamefont {H.}~\bibnamefont {Li}}, \bibinfo {author}
  {\bibfnamefont {W.}~\bibnamefont {Huck}}, \ and\ \bibinfo {author}
  {\bibfnamefont {H.}~\bibnamefont {Sirringhaus}},\ }\href@noop {} {\bibfield
  {journal} {\bibinfo  {journal} {Nat. Mater.}\ }\textbf {\bibinfo {volume}
  {3}},\ \bibinfo {pages} {171} (\bibinfo {year} {2004})}\BibitemShut {NoStop}%
\bibitem [{\citenamefont {Gau}\ \emph {et~al.}(1999)\citenamefont {Gau},
  \citenamefont {Herminghaus}, \citenamefont {Lenz},\ and\ \citenamefont
  {Lipowsky}}]{gau1999liquid}%
  \BibitemOpen
  \bibfield  {author} {\bibinfo {author} {\bibfnamefont {H.}~\bibnamefont
  {Gau}}, \bibinfo {author} {\bibfnamefont {S.}~\bibnamefont {Herminghaus}},
  \bibinfo {author} {\bibfnamefont {P.}~\bibnamefont {Lenz}}, \ and\ \bibinfo
  {author} {\bibfnamefont {R.}~\bibnamefont {Lipowsky}},\ }\href@noop {}
  {\bibfield  {journal} {\bibinfo  {journal} {Science}\ }\textbf {\bibinfo
  {volume} {283}},\ \bibinfo {pages} {46} (\bibinfo {year} {1999})}\BibitemShut
  {NoStop}%
\bibitem [{\citenamefont {Kuang}, \citenamefont {Wang},\ and\ \citenamefont
  {Song}(2014)}]{kuang2014controllable}%
  \BibitemOpen
  \bibfield  {author} {\bibinfo {author} {\bibfnamefont {M.}~\bibnamefont
  {Kuang}}, \bibinfo {author} {\bibfnamefont {L.}~\bibnamefont {Wang}}, \ and\
  \bibinfo {author} {\bibfnamefont {Y.}~\bibnamefont {Song}},\ }\href@noop {}
  {\bibfield  {journal} {\bibinfo  {journal} {Adv. Mater.}\ }\textbf {\bibinfo
  {volume} {26}},\ \bibinfo {pages} {6950} (\bibinfo {year}
  {2014})}\BibitemShut {NoStop}%
\bibitem [{\citenamefont {Hendriks}\ \emph {et~al.}(2008)\citenamefont
  {Hendriks}, \citenamefont {Smith}, \citenamefont {Perelaer}, \citenamefont
  {Van~den Berg},\ and\ \citenamefont {Schubert}}]{hendriks2008invisible}%
  \BibitemOpen
  \bibfield  {author} {\bibinfo {author} {\bibfnamefont {C.~E.}\ \bibnamefont
  {Hendriks}}, \bibinfo {author} {\bibfnamefont {P.~J.}\ \bibnamefont {Smith}},
  \bibinfo {author} {\bibfnamefont {J.}~\bibnamefont {Perelaer}}, \bibinfo
  {author} {\bibfnamefont {A.~M.}\ \bibnamefont {Van~den Berg}}, \ and\
  \bibinfo {author} {\bibfnamefont {U.~S.}\ \bibnamefont {Schubert}},\
  }\href@noop {} {\bibfield  {journal} {\bibinfo  {journal} {Adv. Funct.
  Mater.}\ }\textbf {\bibinfo {volume} {18}},\ \bibinfo {pages} {1031}
  (\bibinfo {year} {2008})}\BibitemShut {NoStop}%
\bibitem [{\citenamefont {Seemann}\ \emph {et~al.}(2005)\citenamefont
  {Seemann}, \citenamefont {Brinkmann}, \citenamefont {Kramer}, \citenamefont
  {Lange},\ and\ \citenamefont {Lipowsky}}]{seemann2005wetting}%
  \BibitemOpen
  \bibfield  {author} {\bibinfo {author} {\bibfnamefont {R.}~\bibnamefont
  {Seemann}}, \bibinfo {author} {\bibfnamefont {M.}~\bibnamefont {Brinkmann}},
  \bibinfo {author} {\bibfnamefont {E.~J.}\ \bibnamefont {Kramer}}, \bibinfo
  {author} {\bibfnamefont {F.~F.}\ \bibnamefont {Lange}}, \ and\ \bibinfo
  {author} {\bibfnamefont {R.}~\bibnamefont {Lipowsky}},\ }\href@noop {}
  {\bibfield  {journal} {\bibinfo  {journal} {Proc. Natl. Acad.Sci.}\ }\textbf
  {\bibinfo {volume} {102}},\ \bibinfo {pages} {1848} (\bibinfo {year}
  {2005})}\BibitemShut {NoStop}%
\bibitem [{\citenamefont {Kant}\ \emph {et~al.}(2017)\citenamefont {Kant},
  \citenamefont {Hazel}, \citenamefont {Dowling}, \citenamefont {Thompson},\
  and\ \citenamefont {Juel}}]{kant}%
  \BibitemOpen
  \bibfield  {author} {\bibinfo {author} {\bibfnamefont {P.}~\bibnamefont
  {Kant}}, \bibinfo {author} {\bibfnamefont {A.~L.}\ \bibnamefont {Hazel}},
  \bibinfo {author} {\bibfnamefont {M.}~\bibnamefont {Dowling}}, \bibinfo
  {author} {\bibfnamefont {A.~B.}\ \bibnamefont {Thompson}}, \ and\ \bibinfo
  {author} {\bibfnamefont {A.}~\bibnamefont {Juel}},\ }\href@noop {} {\bibfield
   {journal} {\bibinfo  {journal} {Phys. Rev. Fluids}\ }\textbf {\bibinfo
  {volume} {2}},\ \bibinfo {pages} {094002} (\bibinfo {year}
  {2017})}\BibitemShut {NoStop}%
\bibitem [{\citenamefont {Kant}\ \emph {et~al.}(2018)\citenamefont {Kant},
  \citenamefont {Hazel}, \citenamefont {Dowling}, \citenamefont {Thompson},\
  and\ \citenamefont {Juel}}]{kant2}%
  \BibitemOpen
  \bibfield  {author} {\bibinfo {author} {\bibfnamefont {P.}~\bibnamefont
  {Kant}}, \bibinfo {author} {\bibfnamefont {A.~L.}\ \bibnamefont {Hazel}},
  \bibinfo {author} {\bibfnamefont {M.}~\bibnamefont {Dowling}}, \bibinfo
  {author} {\bibfnamefont {A.~B.}\ \bibnamefont {Thompson}}, \ and\ \bibinfo
  {author} {\bibfnamefont {A.}~\bibnamefont {Juel}},\ }\href@noop {} {\bibfield
   {journal} {\bibinfo  {journal} {Soft matter}\ }\textbf {\bibinfo {volume}
  {14}},\ \bibinfo {pages} {8709} (\bibinfo {year} {2018})}\BibitemShut
  {NoStop}%
\bibitem [{\citenamefont {Visser}\ \emph {et~al.}(2015)\citenamefont {Visser},
  \citenamefont {Frommhold}, \citenamefont {Wildeman}, \citenamefont {Mettin},
  \citenamefont {Lohse},\ and\ \citenamefont {Sun}}]{visser}%
  \BibitemOpen
  \bibfield  {author} {\bibinfo {author} {\bibfnamefont {C.~W.}\ \bibnamefont
  {Visser}}, \bibinfo {author} {\bibfnamefont {P.~E.}\ \bibnamefont
  {Frommhold}}, \bibinfo {author} {\bibfnamefont {S.}~\bibnamefont {Wildeman}},
  \bibinfo {author} {\bibfnamefont {R.}~\bibnamefont {Mettin}}, \bibinfo
  {author} {\bibfnamefont {D.}~\bibnamefont {Lohse}}, \ and\ \bibinfo {author}
  {\bibfnamefont {C.}~\bibnamefont {Sun}},\ }\href@noop {} {\bibfield
  {journal} {\bibinfo  {journal} {Soft Matter}\ }\textbf {\bibinfo {volume}
  {11}},\ \bibinfo {pages} {1708} (\bibinfo {year} {2015})}\BibitemShut
  {NoStop}%
\bibitem [{\citenamefont {Cox}(1986)}]{cox}%
  \BibitemOpen
  \bibfield  {author} {\bibinfo {author} {\bibfnamefont {R.}~\bibnamefont
  {Cox}},\ }\href@noop {} {\bibfield  {journal} {\bibinfo  {journal} {J. Fluid
  Mech.}\ }\textbf {\bibinfo {volume} {168}},\ \bibinfo {pages} {169} (\bibinfo
  {year} {1986})}\BibitemShut {NoStop}%
\bibitem [{\citenamefont {Voinov}(1976)}]{voinov}%
  \BibitemOpen
  \bibfield  {author} {\bibinfo {author} {\bibfnamefont {O.}~\bibnamefont
  {Voinov}},\ }\href@noop {} {\bibfield  {journal} {\bibinfo  {journal} {Fluid
  Dyn.}\ }\textbf {\bibinfo {volume} {11}},\ \bibinfo {pages} {714} (\bibinfo
  {year} {1976})}\BibitemShut {NoStop}%
\bibitem [{\citenamefont {Heil}\ and\ \citenamefont {Hazel}(2006)}]{oomph}%
  \BibitemOpen
  \bibfield  {author} {\bibinfo {author} {\bibfnamefont {M.}~\bibnamefont
  {Heil}}\ and\ \bibinfo {author} {\bibfnamefont {A.~L.}\ \bibnamefont
  {Hazel}},\ }in\ \href@noop {} {\emph {\bibinfo {booktitle} {Fluid-structure
  interaction}}}\ (\bibinfo  {publisher} {Springer},\ \bibinfo {year} {2006})\
  pp.\ \bibinfo {pages} {19--49}\BibitemShut {NoStop}%
\bibitem [{cod()}]{code}%
  \BibitemOpen
  \href@noop {} {}\bibinfo {howpublished}
  {https://figshare.com/s/9e95e92400ee8e74e6ac}\BibitemShut {NoStop}%
\bibitem [{\citenamefont {Chen}\ \emph {et~al.}(2010)\citenamefont {Chen},
  \citenamefont {Wu}, \citenamefont {Lu},\ and\ \citenamefont {Shieh}}]{LCD}%
  \BibitemOpen
  \bibfield  {author} {\bibinfo {author} {\bibfnamefont {C.-T.}\ \bibnamefont
  {Chen}}, \bibinfo {author} {\bibfnamefont {K.-H.}\ \bibnamefont {Wu}},
  \bibinfo {author} {\bibfnamefont {C.-F.}\ \bibnamefont {Lu}}, \ and\ \bibinfo
  {author} {\bibfnamefont {F.}~\bibnamefont {Shieh}},\ }\href@noop {}
  {\bibfield  {journal} {\bibinfo  {journal} {J. Micromech. Microeng.}\
  }\textbf {\bibinfo {volume} {20}},\ \bibinfo {pages} {055004} (\bibinfo
  {year} {2010})}\BibitemShut {NoStop}%
\end{thebibliography}%

%\noindent \textbf{Supplementary material}

\end{document}